\documentclass{ws-procs9x6}

\newcommand{\gwig}{\mbox{\,\raisebox{.3ex}
    {$>$}$\!\!\!\!\!$\raisebox{-.9ex}{$\sim$}}\,}
\newif\ifhepph
\hepphtrue

\begin{document}

\title{
\ifhepph
\vspace{-3cm}
{\rm\normalsize\rightline{DESY 03-005}\rightline{\lowercase{hep-ph/0301157}}}
\vskip 1cm 
\fi
How to detect the cosmic neutrino background?\ifhepph\footnote{\uppercase{T}alk presented at the
\uppercase{W}orkshop on \uppercase{S}trong and \uppercase{E}lectroweak \uppercase{M}atter 
(\uppercase{SEWM 2002}), \uppercase{O}ctober 2-5, 2002, \uppercase{H}eidelberg, \uppercase{G}ermany.}\fi
         }

\author{A.~Ringwald}

\address{Deutsches Elektronen-Synchrotron DESY,\\
Notkestra\ss e 85, \\ 
D-22607 Hamburg, Germany\\ 
E-mail: andreas.ringwald@desy.de}


\maketitle

\abstracts{
A measurement of the big bang relic neutrinos would open a new window
to the early universe.  We review various possibilities to detect this cosmic neutrino
background  and  substantiate the assertion  that -- apart from the rather indirect 
evidence to be gained  from 
cosmology and large-scale structure formation -- the annihilation of ultrahigh energy
cosmic neutrinos with relic anti-neutrinos (or vice versa) on the $Z$-resonance 
is a  unique process having sensititivy to the relic neutrinos, if a sufficient 
flux at $E^{\rm res}_{\nu_i}=M_Z^2/(2m_{\nu_i} )=4\cdot 10^{22}\ {\rm eV}\,(0.1\ {\rm eV}/m_{\nu_i} )$ 
exists.
The associated absorption dips in the ultrahigh energy
cosmic neutrino spectrum may be searched for at forthcoming neutrino and air shower detectors. The 
associated protons and photons may have been seen already in form of the 
cosmic ray events above the Greisen-Zatsepin-Kuzmin cutoff.   
}

\section{The Cosmic Neutrino Background}

Standard big bang cosmology predicts a diffuse background   
of free photons and neutrinos. The measured cosmic microwave background (CMB)
radiation supports the applicability of standard cosmology
back to photon decoupling which occured approximately three hundred thousand years
after the big bang. The predicted neutrinos from the elusive cosmic neutrino background (C$\nu$B), 
on the other hand,
have decoupled when the universe had a temperature of one MeV and an age
of just one second. Thus, a measurement of the C$\nu$B  
would open a new window to the early universe. 
Its properties are tightly related to the properties of the CMB and are
therefore to be considered as rather firm predictions. In the absence of appreciable
lepton asymmetries one predicts, for example, 
\begin{eqnarray}
\label{av-mom}
\langle |\vec{p}_{\nu_i}|\rangle_0 &=& \langle |{\vec{p}}_{\bar\nu_i}|\rangle_0 
=
3.2\cdot ( 4/11 )^{1/3}\,T_{\gamma\,0} 
=  5\cdot 10^{-4}\ {\rm eV}\,,
\\
\label{av-number}
\langle n_{\nu_i}\rangle_0 
                      & = &  
\langle n_{\bar\nu_i}\rangle_0 
                       =  
(3/22)\, 
\langle n_{\gamma}\rangle_0 
                       =  56\ {\rm cm}^{-3}
\\
\label{omega_nu}
\Omega_{{ {\rm C}\nu {\rm B}}\,0} &=&
2\,\sum_{i=1}^3 m_{\nu_i}\,  
\langle n_{\nu_i}\rangle_0 
/ \rho_c  = (1\cdot 10^{-3}/h^2)\, 
\sum_{i=1}^3  m_{\nu_i}/(0.1\ {\rm eV})\,,
\end{eqnarray}
for todays average 3-momentum $\langle |\vec{p}_{\nu_i}|\rangle_0 $ and number density 
$\langle n_{\nu_i}\rangle_0$ of  
light  ($m_{\nu_i}\ll 1$~MeV) neutrino species $i$, and todays relative contribution 
$\Omega_{{\rm C}\nu {\rm B}\,0}$ to the critical energy density of the universe, in
terms of todays CMB temperature $T_{\gamma\,0}$ and photon number density $\langle n_{\gamma}\rangle_0$. 
The relic neutrino number 
density is comparable to the one of the microwave photons. However,
since neutrinos interact only weakly, the relic neutrinos have not yet been detected directly
in laboratory 
experiments.
Indeed, the average energy of the relic neutrinos is so small, that charged or neutral current 
cross-sections for incoherent scattering off ordinary matter are negligibly small,
\begin{equation}
\label{nuN-cross}
       \sigma_{\nu_i N} \simeq G_F^2\,\langle E_{\nu_i}\rangle^2_0\,/\pi \simeq
       2\cdot 10^{-58}\ {\rm cm}^2\ \left(  m_{\nu_i}/ (0.1\ {\rm eV})\right)^2\,,
\end{equation}
leading to absurdly small event rates, even for kiloton ($N_T\sim 10^{33}$) targets,  
\begin{equation}
\label{rate-incoh}
       R_{\nu_i}^{{\rm ic}} = N_T\,\langle n_{\nu_i}\rangle_0\,\langle |\vec{v}_{\nu_i}|\rangle_0
       \ \sigma_{\nu_i N}
       \simeq 5\cdot 10^{-8}\ {\rm yr}^{-1}\ 
       \left(  N_T/10^{33}\right)\ 
       \left(  m_{\nu_i} / (0.1\ {\rm eV}) \right)  
       .
\end{equation}

Apart from the rather indirect evidence for the C$\nu$B to be gained  from 
cosmology and large-scale structure 
formation\cite{Hu:1997mj
}, which are mainly sensitive to 
$\Omega_{{\rm C}\nu {\rm B}\,0}$~(\ref{omega_nu}), 
two more direct possibilities have been pointed out  in the literature 
and will be outlined in this short review: 
{\em i)} The coherent elastic scattering of the flux of relic neutrinos off target matter in 
a terrestrial detector ({\em flux detection}, Sect.~\ref{flux-det}).  
{\em ii)} The scattering of ultrahigh energy particles (accelerator beams or cosmic rays)
off the relic neutrinos as a target ({\em target detection}, Sect.~\ref{target-det}). 

\begin{figure}[t]
\centerline{
\epsfxsize=3.8cm\epsfbox{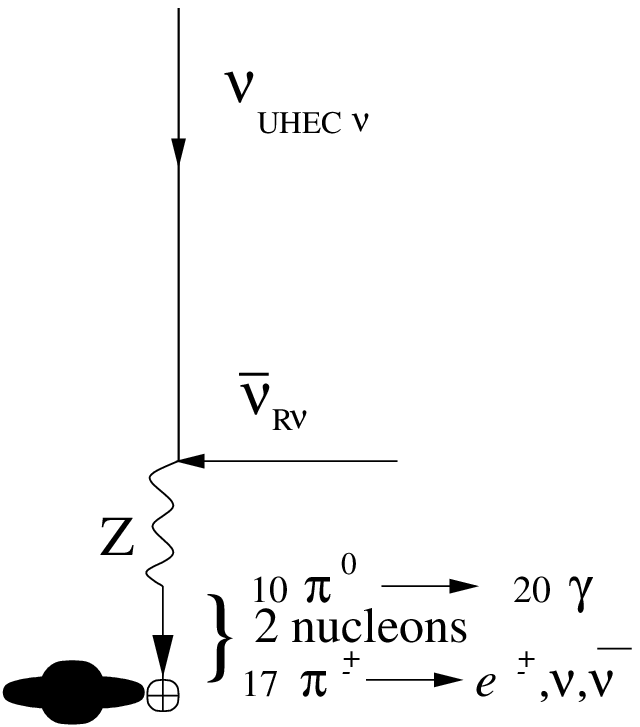}
\hspace{2.3ex}
\epsfxsize=5.7cm\epsfbox{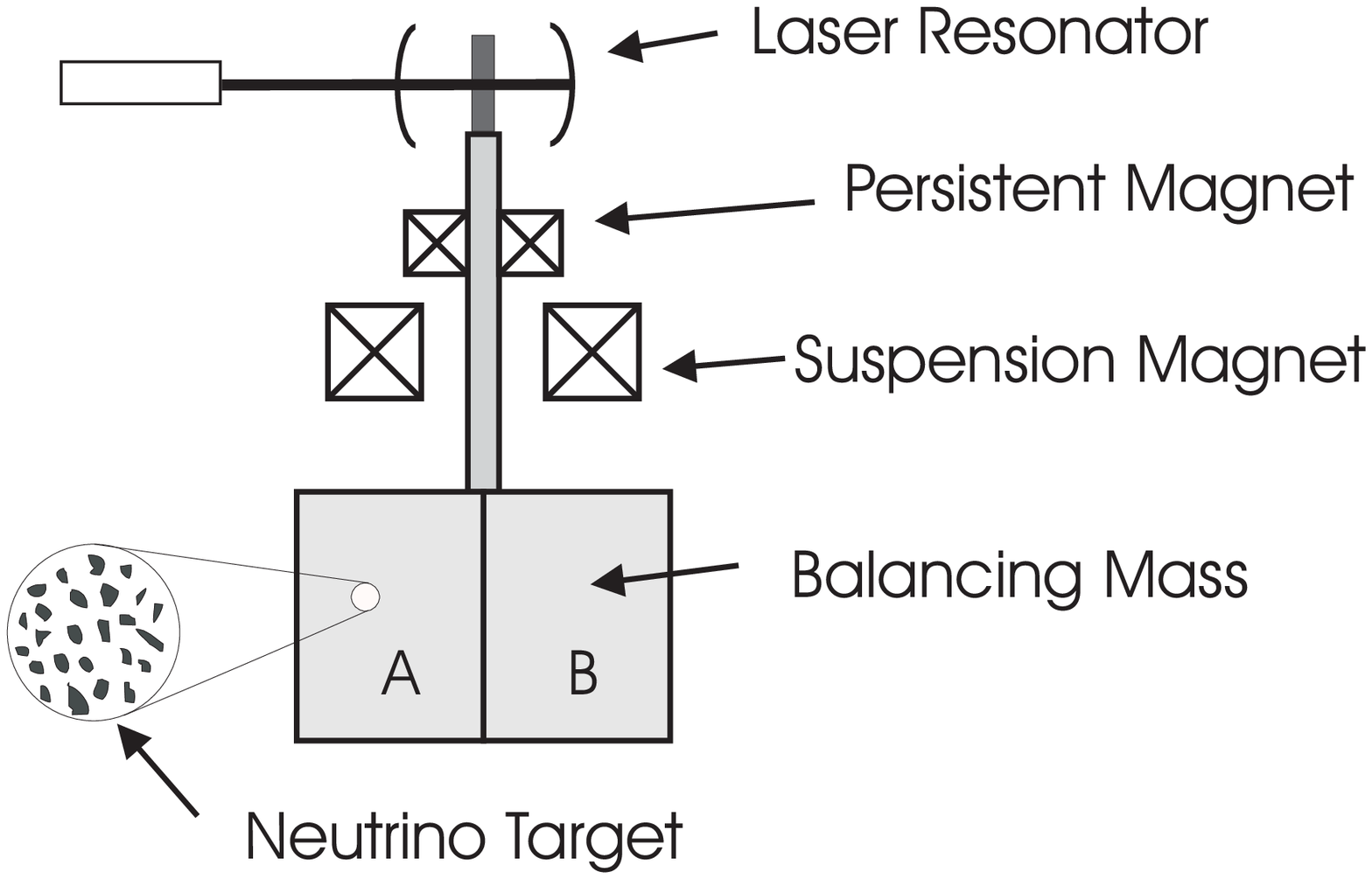}
                   }   
\caption[]{
\label{hagmann-cavendish}
{\em Left:} Annihilation of an ultrahigh energy cosmic neutrino with a 
relic anti-neutrino on the $Z$-resonance (adapted from Ref.\cite{Pas:2001nd}).  
{\em Right:} Schematic diagram of a torsion oscillator proposed to detect the 
relic neutrino wind force\cite{Hagmann:1998nz
}. The target consists
of two hemicylindrical masses with similar densities but different
neutrino cross-section. The target of mass $\sim$ kg is suspended 
by a ``magnetic hook''
consisting of a superconducting magnet in persistent mode
floating above a stationary magnet. The rotation angle is read out with
a tunable optical cavity and an ultra-stable laser.
     }
\end{figure}

Throughout this review, we will take for granted the oscillation interpretation of
atmospheric, solar, and reactor neutrino data\cite{Fukuda:1998mi
}. This, together with the upper mass limit
from tritium $\beta$ decay\cite{Weinheimer:tn
}, implies that the heaviest neutrino has a mass between
${0.04\ {\rm eV}} < { m_{\nu_3}} < { 2.2\ {\rm eV}}$. 
An even stronger -- albeit more model-dependent -- upper bound 
$m_{\nu_3}< 0.8$~eV is obtained from large-scale structure 
formation\cite{Hu:1997mj
}. Such light neutrinos have a very large free streaming length. Therefore, gravitational
clustering of relic neutrinos on the galactic scale can be completely neglected, and 
we base our estimates for terrestrial experiments on the standard cosmological 
value~(\ref{av-number}). Moreover, unclustered, i.e. uniform, enhancements of 
$\langle n_{\nu_i}+n_{\bar\nu_i}\rangle_0$ due to possible neutrino degeneracies  
can also be safely neglected because of recent strong bounds on the latter arising from 
an analysis of big bang nucleosynthesis, taking into account flavor equilibration due to 
neutrino oscillations before $n/p$ 
freeze-out\cite{Lunardini:2000fy
}. Under these conditions, i.e. with no appreciable enhancements of the relic neutrino number densities
in comparison to the standard values~(\ref{av-number}), we shall conclude, in accordance with
Weiler\cite{Weiler:1982qy
}, that the annihilation of ultrahigh energy
cosmic neutrinos with relic anti-neutrinos (or vice versa) on the $Z$-resonance
(cf. Fig.~\ref{hagmann-cavendish} (left)) 
is the  unique process having sensititivy to the relic neutrinos, if a sufficient 
flux at $E^{\rm res}_{\nu_i}=M_Z^2/(2m_{\nu_i} )=4\cdot 10^{22}\ {\rm eV}\,(0.1\ {\rm eV}/m_{\nu_i} )$
exists.       

\section{\label{flux-det} Flux Detection of the C$\nu$B}

The average momentum~(\ref{av-mom}) of relic neutrinos corresponds to a de Broglie wavelength
of macroscopic dimension, 
$\langle \lambda_\nu\rangle_0 = 2\pi/\langle |\vec p_{\nu_i}|\rangle_0 =  0.23$~cm. 
Therefore, one may envisage scattering processes in which many target atoms act 
coherently\cite{Shvartsman:sn
} over a macroscopic volume $\langle \lambda_\nu\rangle_0^3$, so that the reaction rate becomes 
proportional to the square of the number of target atoms in that volume, $N_T^2$, in contrast to 
the incoherent case~(\ref{rate-incoh}). Furthermore, in case of coherent scattering, 
it may be possible to observe the scattering amplitude itself\cite{Stodolsky:1974aq
}, which is linear in $G_F$: ${\mathcal M_{\nu_i}}\sim N_T\,G_F\,m_{\nu_i}$. 
However, in this case one needs a large lepton asymmetry for a non-negligible 
effect. 

A practical scheme to detect the flux of the C$\nu$B by an exploitation of the above coherent
$G_F^2$ effect is based on the fact that a test body of density $\rho_T$ 
at earth will experience a neutrino wind force through random neutrino scattering events, corresponding
to an acceleration\cite{Hagmann:1998nz,
Duda:2001hd}
\begin{eqnarray}
\label{accel}
a_T &= & 
N_A^2\, \rho_T\, \langle \lambda_\nu\rangle_0^3 
\  \langle n_{\nu_i}\rangle_0\,v_{\rm earth}\, 
\sigma_{\nu_i N}\, 
\langle \mid\vec p_\nu\mid\rangle_0 
\\ \nonumber 
&\simeq &
4\cdot 10^{-29}\ {\rm cm}/{\rm s}^{-2}\ 
      ( \rho_T/({\rm g}\,{\rm cm}^{-3}))\,
      ( v_{\rm earth}/(10^{-3}\,c ))\,
      ( m_{\nu_i}/(0.1\ {\rm eV}))^2 
\,,
\end{eqnarray}    
where $N_A$ is Avogadro's constant and $v_{\rm earth}$ is the velocity of the earth 
relative to the CMB. Expression~(\ref{accel}) applies only for 
Dirac neutrinos. For Majorana neutrinos, the acceleration is suppressed 
by a further factor of $(v_{\rm earth}/c)^{2(1)}$ in case of
an unpolarized (polarized) target. 
Therefore, we conclude that this effect is still far from observability. 
At present, the smallest measurable acceleration is $\gwig 10^{-13}$~cm/s$^2$ through 
conventional Cavendish-type torsion balances. Possible improvements to a sensitivity of 
$\gwig 10^{-23}$~cm/s$^2$ have been proposed\cite{Hagmann:1998nz
} (cf. Fig.~\ref{hagmann-cavendish} (right)). However, this is  
still way off the prediction~(\ref{accel}), unless 
one invokes a very unlikely enhancement of the local relic neutrino number density by 
a factor of $10^6$.  

\section{\label{target-det} Target Detection of the C$\nu$B}

Let us consider next the idea to take advantage of the fact that 
at center-of-mass (cm) energies below the $W$- and $Z$-resonances the neutrino cross-sections
are rapidly growing with energy. Correspondingly, one may envisage the possibility to 
exploit a flux of ultrahigh energy particles -- either from accelerator beams or from
cosmic rays -- for scattering on the C$\nu$B. However, the attainable cm energies, 
\begin{equation}
\sqrt{s}=  \sqrt{2\,m_\nu\,E_{\rm beam}}= 0.4\ {\rm MeV}\  
( m_\nu/(0.1\ {\rm eV}))^{1/2}\,
( E_{\rm beam}/(1\ {\rm TeV}))^{1/2},
\end{equation}
at forthcoming 
accelerator beams such as TESLA/LHC/VLHC, with beam energies $E_{\rm beam}$ of 0.5/7/100 TeV, 
are so low,  
that the cross-sections for such interactions are still quite small,
\begin{equation}
       {\sigma_{\nu\,{\rm beam}}} \simeq G_F^2\,s\,/\pi \simeq  
       {3\cdot 10^{-46}\ {\rm cm}^2}\ 
( m_{\nu_i}/(0.1\ {\rm eV}))\,
( E_{\rm beam}/(1\ {\rm TeV})))
\,, 
\end{equation}
leading to a terribly small scattering rate of\cite{Muller:1987qm
}
\begin{equation}
       { R_{\nu_i\,{\rm beam}}} \simeq  
       {4\cdot 10^{-12}\ {\rm yr}^{-1}}\ 
       \left( \frac{I}{\rm A}\right)\,\left( \frac{L}{10\ {\rm km}}\right)\,
       \left( \frac{m_{\nu_i}}{0.1\ {\rm eV}}\right)\,
       \left( \frac{E_{\rm beam}}{1\ {\rm TeV}}\right)\,
\,,
\end{equation}
for a beam of length $L$ and current $I$.  
Thus, there is little hope for detection of the C$\nu$B using terrestrial
accelerator beams in the foreseeable future.     

\begin{figure}[t]
\vspace{-1.2cm}
\centerline{\epsfxsize=5.8cm\epsfbox{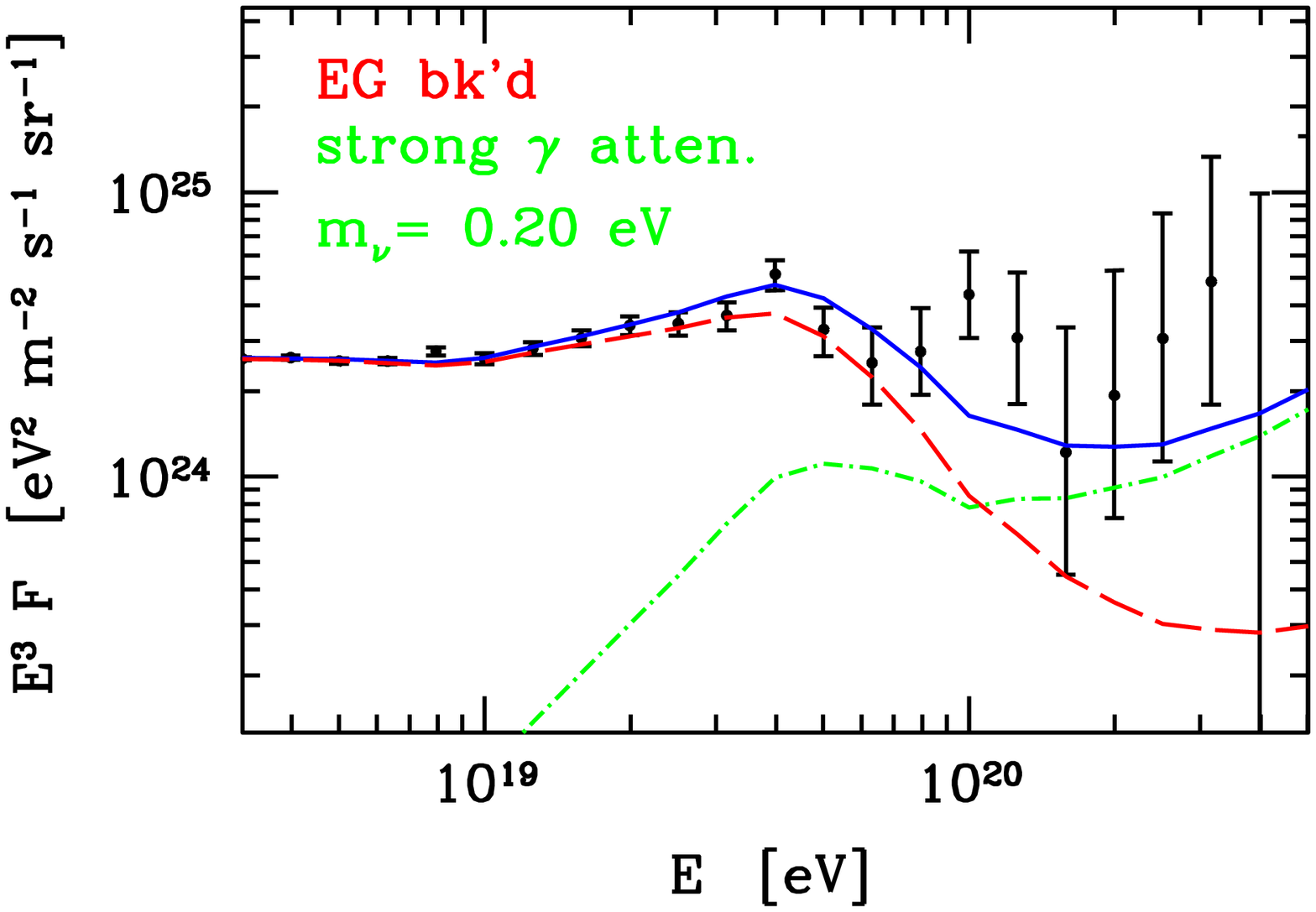}
\hspace{-1ex}
                   \epsfxsize=5.8cm\epsfbox{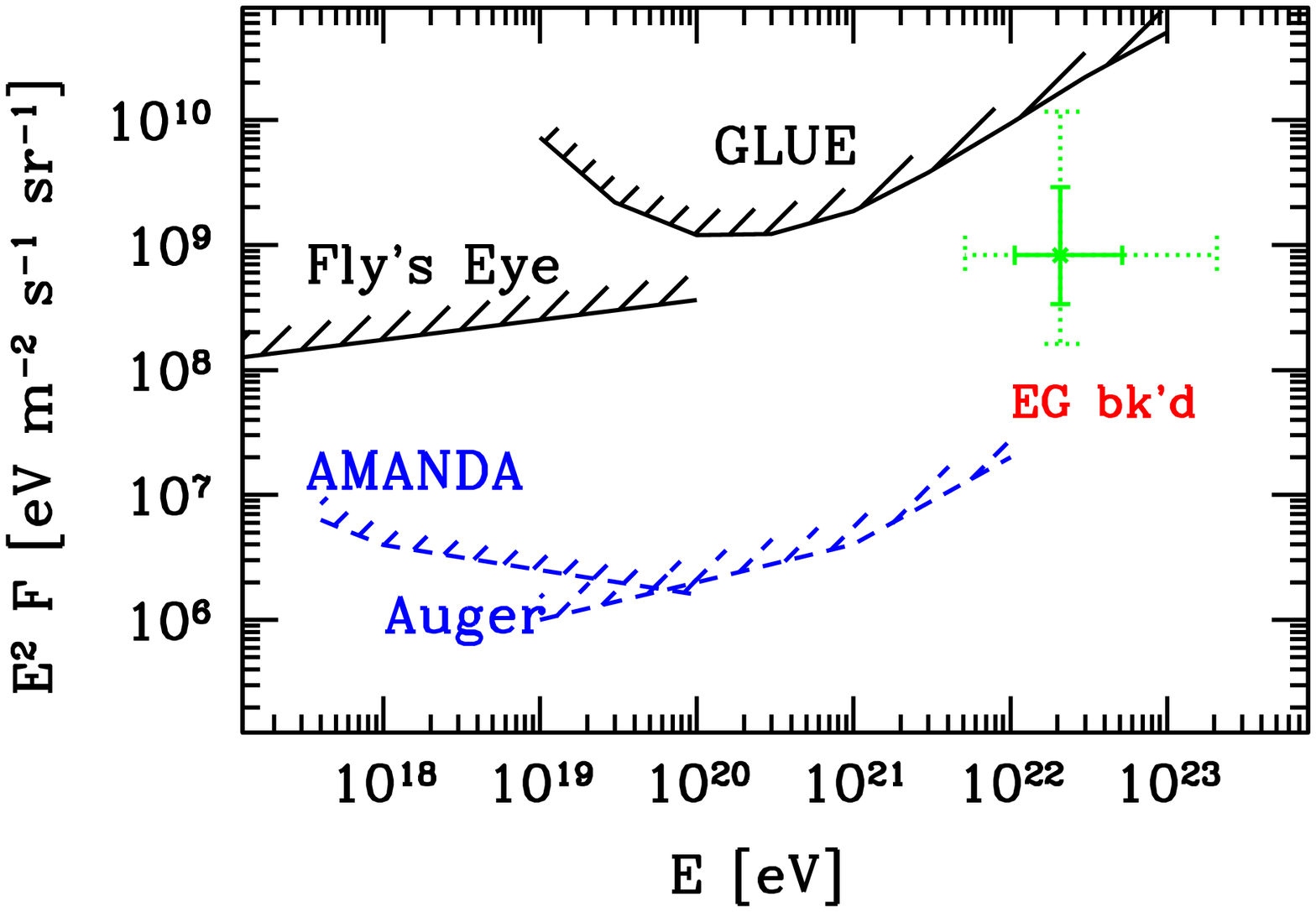}
                    }   
\vspace{-1.cm}
\caption[]{{\em Left:} Observed ultrahigh energy cosmic ray 
spectrum\cite{Takeda:1998ps,Bird:yi,
Lawrence:cc,
Kieda00
} (points with error bars),  
in comparison to the predicted one  in the $Z$-burst scenario\cite{Fodor:2001qy
} (solid), originating from  a background 
of ordinary cosmic ray nucleons of extragalactic origin (dashed) plus nucleons 
from e.g. $\nu_{{\rm UHEC}\nu}+\bar\nu_{{\rm C}\nu{\rm B}}\to Z\to N\bar N+X$ 
(dashed-dotted). 
{\em Right:} Upper limits on the ultrahigh energy cosmic neutrino flux from the Fly's Eye and 
Goldstone Lunar Experiment (shaded-solid) and projected upper limits from
AMANDA and the Pierre Auger Observatory (shaded-dashed),  in comparison
to the prediction in the $Z$-burst scenario\cite{Fodor:2001qy
} (point with error bars).  
\label{uhecr-spectrum}}
\end{figure}

Let us finally consider cosmic rays. Ultrahigh energy cosmic rays have 
been seen by air shower observatories such as AGASA\cite{Takeda:1998ps}, 
Fly's Eye\cite{Bird:yi
}, Haverah Park\cite{Lawrence:cc
}, HiRes\cite{Kieda00
}, and Yakutsk\cite{Efimov91}, 
up to energies $E_{\rm cr}\sim 10^{20}$~eV, corresponding to cm energies
\begin{equation}
\sqrt{s}=  \sqrt{2\,m_\nu\,E_{\rm beam}}= 4\ {\rm GeV}\  
( m_\nu/(0.1\ {\rm eV}))^{1/2}\,
( E_{\rm cr}/(10^{20}\ {\rm eV}))^{1/2}. 
\end{equation}
The latter are not too far away from the $W$- and $Z$-resonances, at which the 
electroweak cross-sections get sizeable. Indeed, it has been pointed out
long ago by Weiler\cite{Weiler:1982qy
}, that the resonant annihilation of ultrahigh energy cosmic neutrinos 
with relic (anti-)neutrinos on the $Z$-boson appears to be a unique  
process\ifhepph\footnote{For earlier and related suggestions, see 
Ref.\cite{Bernstein:1963
} and Ref.\cite{Wigmans:2002rb
}, respectively.}\fi\   
having sensitivity to the C$\nu$B. 
On resonance, $E^{\rm res}_\nu=M_Z^2/(2m_\nu )=4\cdot 10^{22}\ {\rm eV}\,(0.1\ {\rm eV}/m_\nu )$, 
the corresponding cross-section is enhanced by several orders of magnitudes, 
\begin{equation}
\langle\sigma_{\rm ann}\rangle =\int {\rm d}s/M_Z^2\ \sigma_{\nu\bar\nu}^Z(s)=
2\pi\sqrt{2}\,G_F\simeq 4\cdot 10^{-32}\ {\rm cm}^2
,
\end{equation} 
leading to a ``short'' mean free path $\ell_{\nu_i\,0}=(\langle n_{\nu_i}\rangle_0\langle \sigma_{\rm ann}
\rangle)^{-1}\simeq 1.4\cdot 10^5$~Mpc which is ``only'' about $48\,h$ times the Hubble distance. This 
corresponds to an annihilation probability for ultrahigh energy neutrinos from cosmological distances 
on the C$\nu$B of $2\,h^{-1}$~\%, neglecting cosmic evolution effects. 
The signatures of annihilation might be {\em i)} absorption 
dips\cite{Weiler:1982qy,
Roulet:1993pz
} in the ultrahigh energy
cosmic neutrino spectrum at the resonant energies and {\em ii)} emission 
features\cite{Fargion:1999ft
} ($Z$-bursts) 
as protons (or photons) (cf. Fig.~\ref{hagmann-cavendish} (left)) above the predicted 
Greisen-Zatsepin-Kuzmin-cutoff\cite{Greisen:1966jv
} at 
$E_{\rm GZK}\simeq 4\cdot 10^{19}$~eV. 

In fact, since Weiler's 1982 proposal of absorption dips, a (significant(?))  
number of cosmic rays with energies above $E_{\rm GZK}$ has been accumulated by air 
shower observatories\cite{Takeda:1998ps,Bird:yi,
Lawrence:cc,
Kieda00,
Efimov91} (cf. Fig.~\ref{uhecr-spectrum} (left)). 
This presents a puzzle, since these cosmic rays of most probably extragalactic 
origin\ifhepph\footnote{Plausible astrophysical
sources for those energetic particles are at cosmological distances.}\fi\ 
should show a pronounced depletion above 
$E_{\rm GZK}$ (cf. Fig.~\ref{uhecr-spectrum} (left)), since nucleons with super-GZK energies have a 
short energy attenuation length of about $50$~Mpc due
to inelastic interactions with the CMB. 
Ultrahigh energy neutrinos produced at cosmological distances, on the other hand, can reach the
GZK zone unattenuated and their resonant annihilation on the relic neutrinos could just result 
in the observed cosmic rays beyond $E_{\rm GZK}$. 

The energy spectrum of the highest energy cosmic rays depends
critically on the neutrino mass if they are indeed produced via 
$Z$-bursts\cite{Fargion:1999ft,
Pas:2001nd,Fodor:2001qy,
Gelmini:2002xy
}. From a quantitative comparison of the predicted spectrum
with the observed one (cf. Fig.~\ref{uhecr-spectrum} (left)), one can therefore infer the required 
mass of the heaviest neutrino\cite{Fodor:2001qy
}. The value of the neutrino mass obtained in this way
is fairly robust against variations in presently unknown quantities, such as
the amount of the universal radio background and the
extragalactic magnetic field, within their anticipated uncertainties. 
It  turns out to lie in the range 
$0.08\ {\rm eV} \leq m_{\nu_3}\leq 1.3$~eV at the 68\,\% confidence level, 
which compares favourably with the present 
knowledge coming from oscillations, tritium beta decay\cite{Weinheimer:tn
}, and neutrinoless
double beta decay\cite{Klapdor-Kleingrothaus:2001yx
}. 
This range narrows down considerably if a particular universal radio background is assumed, e.g. 
to $0.08\ {\rm eV}\leq m_{\nu_3}\leq 0.40$~eV for a large one. 

The required ultrahigh energy cosmic neutrino fluxes (cf. Fig.~\ref{uhecr-spectrum} (right)) 
should be observed in the near future by existing neutrino telescopes,  
such as AMANDA and RICE, and by cosmic ray air shower detectors
currently under construction,   
such as the Pierre Auger Observatory.
\ifhepph Otherwise the $Z$-burst scenario
for the origin of the highest-energy cosmic rays
will be ruled out.
The required neutrino fluxes are enormous.\fi\  
If such tremendous fluxes of ultrahigh energy neutrinos 
are indeed found, one has to deal with the challenge to explain their origin.
It is fair to say, that at the moment no convincing astrophysical
sources are known which meet the requirements of the $Z$-burst scenario, i.e. which  
accelerate protons at least up to $10^{23}$~eV,  
are opaque to primary
nucleons, and emit secondary photons only in the sub-MeV region\cite{Kalashev:2001sh
}. 
However, even if the ultrahigh energy cosmic neutrino flux turns out to be too small for the above
$Z$-burst scenario to be realized, 
a far future precision search for absorption dips in the resonant region\ifhepph\footnote{Assuming 
that $m_{\nu_3}$ is then already known from laboratory 
experiments.}\fi\   
-- presumably beyond the sensitivity of e.g. the projected 
Extreme Universe Space Observatory (EUSO) -- 
may still reveal the existence of the C$\nu$B.  

\section*{Acknowledgments}

I would like to thank Z.~Fodor and S.~D.~Katz for the nice collaboration \ifhepph on the $Z$-burst scenario 
and a careful reading of the manuscript\fi
.

\end{document}